\documentclass[12pt]{article}
\usepackage{amsmath,amsfonts,amssymb,picture,graphicx}

\begin{document}
\thispagestyle{empty}

\def\theequation{\arabic{section}.\arabic{equation}}
\def\a{\alpha}
\def\b{\beta}
\def\g{\gamma}
\def\d{\delta}
\def\dd{\rm d}
\def\e{\epsilon}
\def\ve{\varepsilon}
\def\z{\zeta}
\def\teta{\tilde\eta}

\newcommand{\h}{\hspace{0.5cm}}

\begin{titlepage}
\vspace*{-1.cm}
\renewcommand{\thefootnote}{\fnsymbol{footnote}}
\begin{center}
{\Large \bf NLIE for the Sausage model}
\end{center}
\vskip .8cm \centerline{\bf Changrim  Ahn$^1$, Janos Balog$^2$, and Francesco Ravanini$^{3,4}$ }

\vskip 10mm

\centerline{\sl $^1$Department of Physics} \centerline{\sl Ewha Womans University}
 \centerline{\sl DaeHyun 11-1, Seoul 120-750, S. Korea}

\vskip 0.8cm

\centerline{\sl $^2$Institute for Particle and Nuclear Physics}
\centerline{\sl Wigner Research Centre for Physics, MTA Lend\"ulet Holographic QFT Group} 
\centerline{\sl 1525 Budapest 114, P.O.B. 49, Hungary}

\vskip 0.8cm

\centerline{\sl $^3$Department of Physics and Astronomy}
\centerline{\sl University of Bologna}
\centerline{\sl Via Irnerio 46, 40126 Bologna, Italy}

\vskip 0.3cm

\centerline{\sl $^4$ Istituto Nazionale di Fisica Nucleare, Sezione di Bologna}
\centerline{\sl Via Irnerio 46, 40126 Bologna, Italy}

\vskip 10mm

\baselineskip 18pt

\begin{center}
{\bf Abstract}
\end{center}
The sausage model, first proposed by Fateev, Onofri, and Zamolodchikov, is a deformation of the $O(3)$ sigma model preserving integrability.
The target space is deformed from the sphere to ``sausage'' shape by a deformation parameter $\nu$.
This model is defined by a factorizable $S$-matrix which is 
obtained by deforming that of the $O(3)$ sigma model by a parameter $\lambda$.
Clues for the deformed sigma model are provided by various UV and IR information through 
the thermodynamic Bethe ansatz (TBA) analysis based on the $S$-matrix.
Application of TBA to the sausage model is, however,  limited to the case of $1/\lambda$ integer where the coupled integral equations can be truncated to a finite number.
In this paper, we propose a finite set of nonliear integral equations (NLIEs), which are applicable to generic value of $\lambda$.
Our derivation is based on $T-Q$ relations extracted from the truncated TBA equations.
For consistency check, 
we compute next-leading order corrections of the vacuum energy and extract the $S$-matrix information in the 
IR limit.
We also solved the NLIE both analytically and numerically in the UV limit to get the effective central charge and compared with
that of the zero-mode dynamics to obtain exact relation between $\nu$ and $\lambda$. 
{\it This paper is a tribute to the memory of Prof. Petr Kulish}.

\end{titlepage}
%\end{quote}
%\vskip 1cm \centerline{\today}
\newpage
\baselineskip 18pt

%%%%%%%%%%%%%%%%%%%%%%%%%%%%%%%%%%%%%%%%%%%%%%%%%
\def\nn{\nonumber}
%%%%%%%%%%%%%%%%%%%%%%%%%%%%%%%%%%%%%%%%%%%%%%%%%
%%%%%%%%%%%%%%%%%%%%%%%%%%%%%%%%%%%%%%%%%%%%%%%%%%%%%
\def\tr{{\rm tr}\,}
\def\p{\partial}
\newcommand{\non}{\nonumber}
\newcommand{\bea}{\begin{eqnarray}}
\newcommand{\eea}{\end{eqnarray}}
\newcommand{\bde}{{\bf e}}
\renewcommand{\thefootnote}{\fnsymbol{footnote}}
\newcommand{\be}{\begin{eqnarray}}
\newcommand{\ee}{\end{eqnarray}}
\newcommand{\beq}{\begin{equation}}
\newcommand{\eeq}{\end{equation}}
%\newcommand{\h}{\hspace{0.5cm}}
%%%%%%%%%%%%%%%%%%%%%%%%%%%%%%%%%%%%%%%%%%%%%%%%%%%%

\vskip 0cm

\renewcommand{\thefootnote}{\arabic{footnote}}
\setcounter{footnote}{0}

\setcounter{equation}{0}
%%%%%%%%%%%%%5%%%%%%%%%%%%%%%%%%%%%%%%%%%%%%%%%%%%%%
\section{Introduction}

Two-dimensional nonlinear sigma (NLS) models form an interesting class of quantum field theories as they may
describe string theories on nontrivial target manifolds, continuum spin systems, quantum gravity and black holes.
Even more interesting subclasses of NLS models are those which can be exactly solvable. 
These provide valuable information on non-perturbative aspects of quantum fields.
One of the many recent applications of these models appears in the AdS/CFT correspondence \cite{Malda} which is largely based on the integrability discovered in the target space \cite{review}.
Thanks to integrability, the $S$-matrix is factorizable and can be applied to compute finite-size effects of the 
NLS models.
Thermodynamic Bethe ansatz (TBA), which is directly derived from the $S$-matrix, is a most
commonly used method for this purpose \cite{tba}.
While it is an efficient tool for some class of integrable models, the TBA gets complicated for the NLS models which typically introduce an infinite set of coupled integral equations.
To overcome this technical problem, nonlinear integral equations (NLIEs) are constructed 
for the finite-size effects \cite{nlie1, nlie2,nlie3} which replace the infinite number of TBA equations with only 
a finite one.
There is a disadvantage, however, that the connections between the NLIEs and the $S$-matrix of the orginal model are more involved.
This can be overcome if the NLIEs can be derived from the TBAs.
These derivations are available for various 2d NLS models \cite{BalHeg} and 
for the AdS/CFT \cite{adsnlie,Gromov:2011cx}.

Another direction of developments in the study of the NLS models is to extend the target spaces which preserve
integrability. 
The sausage model is one of the earliest attempts in this direction \cite{FOZ}.
Fateev, Onofri, and Zamolodchikov have considered a deformation of the $O(3)$ NLS model 
which can still be integrable.
The target space is deformed from the sphere to ``sausage'' shape by a deformation parameter $\nu$.
Assuming the integrability rather than proving it, the authors have proposed exact $S$-matrix by deforming
that of $O(3)$ model with a parameter $\lambda$ and have computed various physical quantities, such as
finite-size effects.
This kind of generalization has been also studied in the AdS/CFT recently under the names of
$\gamma$-, $\eta$-, and $\kappa$-deformations \cite{adsdeform1, adsdeform2, adsdeform3, adsdeform4} whose $S$-matrices on the worldsheet are deformed by such parameters \cite{BeiKor} while preserving integrability.

Introducing the deformation parameter $\lambda$ raises various technical issues.
A new set of bound-states can appear for certain domain of $\lambda$ which complicates the TBA equations further.
Analytic relationship between the $\lambda$ of the $S$-matrix and target space deformation $\nu$ should be 
necessary for the complete understanding.
While particle spectrum of the sausage model remains simple in the domain of $0\le\lambda<1/2$, the 
TBA has been derived only for integer values of $1/\lambda$ \cite{FOZ}.
The main goal of this paper is to derive NLIE equations applicable to generic values of $\lambda$,
but still limited to the above domain.
Our derivation is based on the manipulation of the TBA system, or equivalently its functional equation form, the so-called 
``$Y$-system'' along with analytic
properties in the line of direct derivation of NLIEs from the TBA \cite{BalHeg}.
Since the TBA has been constructed for the integer values of $1/\lambda$, an analytic continuation to non-integer
$1/\lambda$ should be
assumed at certain step.
The validity of this assumption is checked a posteriori by deriving $S$-matrix elements from the NLIE in the large volume limit.
In the opposite short distance or UV limit, we can solve the NLIE equations either analytically or numerically and
find an exact relation between $\lambda$ and $\nu$,
which turns out to be different from the one conjectured in the original paper \cite{FOZ}.

This paper is organized as follows. In sect.2 we summarize the relevant contents of the sausage model in \cite{FOZ}.
Sect.3 contains our main results. We derive the NLIE equations and analyze both IR and UV limits.
We conclude this paper in sect.4 with brief summary and possible open problems.
We explain the details of analytic UV computations in Appendix A.

\setcounter{equation}{0}
\section{Sausage sigma model as factorized $S$-matrix theory}

The $O(3)$ NLS model is a prototype of an integrable model with action
\be
{\cal A}_{O(3)}=\frac{1}{2g}\int \sum_{a=1}^3(\partial_\mu n_a)^2 d^2 x+i\theta T
\label{O3action}
\ee
where $T$ is a Wess-Zumino topological term. 
The three fields $n_a$ are an $O(3)$ unit vector $\sum_an_a^2=1$.
This model, denoted by ${\rm SSM}_0^{(\theta)}$ \cite{FOZ}, is integrable for $\theta=0,\pi$.
Although the two cases show the same UV behaviour, they are very different in IR.
The particle spectrum of ${\rm SSM}_0^{(0)}$ is a massive triplet of $O(3)$, whose $S$-matrix is $O(3)$-invariant\cite{ZamZam1},
\be
{\mathbf S}(\theta)&=&S_0(\theta){\mathbf P}_0+S_1(\theta){\mathbf P}_1+S_2(\theta){\mathbf P}_2
\label{O3massive}\\
S_0(\theta)&=&\frac{\theta+2i\pi}{\theta-2i\pi},\quad S_1(\theta)=\frac{(\theta-i\pi)(\theta+2i\pi)}{(\theta+i\pi)(\theta-2i\pi)},\quad
S_2(\theta)=\frac{\theta-i\pi}{\theta+i\pi},
\label{O3massivei}
\ee
where ${\mathbf P}_j,\ j=0,1,2$ are projectors on the $j$-spin states.
The ${\rm SSM}_0^{(\pi)}$ sigma model instead interpolates the UV CFT to an IR CFT which is a WZW $SU(2)_1$ model.
The spectrum consists of two doublets, left(L)- and right(R)-moving.
$L-L$, $R-R$, and $R-L$ scattering matrices are all the same and given by \cite{ZamZam2} 
\be
S^{(LL)}(\theta)=S^{(RR)}(\theta)=S^{(LR)}(\theta)=\frac{\Gamma\left(\frac{1}{2}+\frac{\theta}{2i\pi}\right)\Gamma\left(-\frac{\theta}{2i\pi}\right)}
{\Gamma\left(\frac{1}{2}-\frac{\theta}{2i\pi}\right)\Gamma\left(\frac{\theta}{2i\pi}\right)}\frac{\theta{\mathbf 1}-i\pi{\cal P}}{\theta-i\pi}
\label{O3massless}
\ee
with the permutation matrix ${\cal P}$.

The sausage model is defined by a deformation of the above $S$-matrices.
These scattering theories, denoted by ${\rm SST}_\lambda^{(\pm)}$, have the same particle spectrum as ${\rm SSM}_0^{(0,\pi)}$,
respectively.
The non-vanishing $S$-matrix elements of ${\rm SST}_\lambda^{(+)}$ for the triplet $(-,0,+)$ are \cite{FOZ}
\be
S_{++}^{++}(\theta)&=&S_{+-}^{+-}(i\pi-\theta)=\frac{\sinh\left(\lambda(\theta-i\pi)\right)}{\sinh\left(\lambda(\theta+i\pi)\right)},
\label{SSTpp}\\
S_{+0}^{0+}(\theta)&=&S_{+-}^{00}(i\pi-\theta)=\frac{-i\sin(2\pi\lambda)}{\sinh\left(\lambda(\theta-2i\pi)\right)}S_{++}^{++}(\theta),\\
S_{+0}^{+0}(\theta)&=&\frac{\sinh\left(\lambda\theta\right)}{\sinh\left(\lambda(\theta-2i\pi)\right)}S_{++}^{++}(\theta),\\
S_{+-}^{-+}(\theta)&=&-\frac{\sin(\pi\lambda)\sin(2\pi\lambda)}{\sinh\left(\lambda(\theta-2i\pi)\right)\sinh\left(\lambda(\theta+i\pi)\right)},\quad
S_{00}^{00}(\theta)=S_{+0}^{+0}(\theta)+S_{-+}^{+-}(\theta).
\label{SSTp}
\ee
This $S$-matrix reduces to (\ref{O3massive}) and (\ref{O3massivei}) in the $\lambda\to 0$ limit.
If $0\le \lambda<1/2$, all the $S$-matrix elements in (\ref{SSTpp})-(\ref{SSTp}) have no poles in the physical strip 
$0\le \Im m\ \theta<\pi$. 
At $\lambda=1/2$ the theory becomes free and the triplet becomes a complex fermion and a boson with the same mass.
The ${\rm SST}_\lambda^{(+)}$ becomes very complicated in the domain of $\lambda>1/2$.
The $S$-matrix elements have bound-state poles which should be analyzed by complete bootstrap processes
and there is no evidence that the scattering theory is a consistent one.
We focus only in the ``repulsive'' domain $0\le \lambda< 1/2$ in this paper.

The non-vanishing $S$-matrix elements of ${\rm SST}_\lambda^{(-)}$ between two ($L$- and $R$-movers) set of massless doublets $(+,-)$ 
are given by \cite{FOZ}
\be
U_{++}^{++}(\theta)&=&U_{--}^{--}(\theta)=U_{0}(\theta),\\
U_{+-}^{+-}(\theta)&=&U_{-+}^{-+}(\theta)=-\frac{\sinh\left(\lambda\theta/(1-\lambda)\right)}{\sinh\left(\lambda(\theta-i\pi)/(1-\lambda)\right)}U_{0}(\theta),\\
U_{-+}^{+-}(\theta)&=&U_{+-}^{-+}(\theta)=-i\frac{\sin\left(\pi\lambda/(1-\lambda)\right)}{\sinh\left(\lambda(\theta-i\pi)/(1-\lambda)\right)}
U_{0}(\theta),\\
U_{0}(\theta)&=&-\exp\left[i\int_0^{\infty}\frac{\sinh\left((1-2\lambda)\pi\omega/(2\lambda)\right)\sin(\omega\theta)}
{\cosh(\pi\omega/2)\sinh\left((1-\lambda)\pi\omega/(2\lambda)\right)}\frac{d\omega}{\omega}\right].
\label{SSTm}
\ee
As $\lambda\to 0$, this reduces to (\ref{O3massless}).

Main claim of \cite{FOZ} is that the scattering theories ${\rm SST}_\lambda^{(\pm)}$ correspond to a deformed sigma model
described by an effective action with $\theta=0,\pi$
\be
{\cal A}_{{\rm SSM}^{(\theta)}_\nu}=\int \frac{(\partial_\mu Y)^2+(\partial_\mu X)^2}
{a(t)+b(t)\cosh(2Y)}\  d^2 x+i\theta T,
\label{SSMaction}
\ee
with RG flows in the leading order given by 
\be
a(t)=-\nu\coth\left(\frac{\nu(t-t_0)}{2\pi}\right),\qquad b(t)=-\nu/\sinh\left(\frac{\nu(t-t_0)}{2\pi}\right).
\ee
By comparing bulk free energy from this action coupled with an external field with Bethe ansatz computation based on the $S$-matrix, 
the authors of \cite{FOZ} have found the relation in the weak coupling region,
\be
\nu=4\pi\lambda+{\cal O}(\lambda^2).
\label{weaknulam}
\ee

Another important support comes from the thermodynamic Bethe ansatz (TBA) analysis.
In the UV limit $t\sim -\infty$, where the target space looks like a long sausage with a length $L$ and a circumference $l$
\be
L\approx \frac{\sqrt{2\nu}}{2\pi} (t_0-t),\qquad
l=2\pi\sqrt{\frac{2}{\nu}},
\label{sizes}
\ee
one can compute the effective central charge from the Schr\"odinger equation of the zero-mode of the field $Y$ 
based on the effective action (\ref{SSMaction}) which is valid in semi-classic limit $\nu\ll 1$.
The central charge is expressed as a function of the system size $r$ which is related to the RG scale by
$t-t_0=\log(r\Lambda_0)$;
\be
c_{\nu}(r)=2-\frac{\nu}{4\pi}\left[\frac{3\pi^2}{2(\eta+2\log 2)^2}+{\cal O}(\eta^{-4})\right],\qquad{\rm with}\quad
\eta=\frac{\nu}{4\pi}(t_0-t).
\label{ccSSM}
\ee

In the factorizable scattering theory side, the TBA can be used for the effective central charge.
Derivation of the TBA for the ${\rm SST}_\lambda^{(\pm)}$ is not trivial, however, due to the matrix structure of the $S$-matrix.
A direct derivation is viable only for a special value of $\lambda$, namely,
\be
\lambda=\frac{1}{N},\qquad N=2,3,\ldots.
\ee
For this case, the TBA system includes only finite number of unknown functions $\varepsilon_a$ ($a=0,1,\ldots,N$) which
satisfy
\be
r\rho_a(\theta)=\varepsilon_a(\theta)+\sum_{b=0}^N\int_{-\infty}^{\infty}\frac{l_{ab}}{\cosh(\theta-\theta')}\log\left(1+e^{-\varepsilon_b(\theta')}\right)\frac{d\theta'}{2\pi},
\label{TBA}
\ee
where the driving terms are $\rho_a(\theta)$ are 
\be
\rho_a(\theta)=m\delta_{a0}\cosh\theta\quad{\rm for}\quad {\rm SST}_\lambda^{(+)};\qquad
\rho_a(\theta)=\frac{m}{2}\left(\delta_{a0}e^{\theta}+\delta_{a1}e^{-\theta}\right)\quad{\rm for}\quad {\rm SST}_\lambda^{(-)},
\ee
and $l_{ab}$ is the incidence matrix of the graph given in Fig.1.
\begin{figure}
\begin{picture}(400,200)(-40,0)
\put(20,80){\circle*{10}}
\put(20,20){\circle*{10}}
\put(50,50){\circle{10}}
\put(55,50){\line(1,0){35}}
\put(95,50){\circle{10}}
\put(100,50){\line(1,0){35}}
\put(140,50){\circle{10}}
\put(145,50){\line(1,0){20}}
\put(23,23){\line(1,1){24}}
\put(23,77){\line(1,-1){24}}
\multiput(172,50)(10,0){6}{\line(1,0){5}}
\put(380,80){\circle{10}}
\put(380,20){\circle{10}}
\put(350,50){\circle{10}}
\put(345,50){\line(-1,0){35}}
\put(305,50){\circle{10}}
\put(300,50){\line(-1,0){35}}
\put(260,50){\circle{10}}
\put(255,50){\line(-1,0){20}}
\put(377,23){\line(-1,1){24}}
\put(377,77){\line(-1,-1){24}}
%%%
\put(16,90){$1$}
\put(16,30){$0$}
\put(46,60){$2$}
\put(91,60){$3$}
\put(136,60){$4$}
\put(374,90){$N$}
\put(374,30){$N-1$}
\put(324,60){$N-2$}
\put(279,60){$N-3$}
\put(234,60){$N-4$}
\put(-15,80){$\frac{m}{2}e^{-\theta}$}
\put(-10,20){$\frac{m}{2}e^{\theta}$}
\put(20,180){\circle{10}}
\put(20,120){\circle*{10}}
\put(50,150){\circle{10}}
\put(55,150){\line(1,0){35}}
\put(95,150){\circle{10}}
\put(100,150){\line(1,0){35}}
\put(140,150){\circle{10}}
\put(145,150){\line(1,0){20}}
\put(23,123){\line(1,1){24}}
\put(23,177){\line(1,-1){24}}
\multiput(172,150)(10,0){6}{\line(1,0){5}}
\put(380,180){\circle{10}}
\put(380,120){\circle{10}}
\put(350,150){\circle{10}}
\put(345,150){\line(-1,0){35}}
\put(305,150){\circle{10}}
\put(300,150){\line(-1,0){35}}
\put(260,150){\circle{10}}
\put(255,150){\line(-1,0){20}}
\put(377,123){\line(-1,1){24}}
\put(377,177){\line(-1,-1){24}}
%%%
\put(16,190){$1$}
\put(16,130){$0$}
\put(46,160){$2$}
\put(91,160){$3$}
\put(136,160){$4$}
\put(374,190){$N$}
\put(374,130){$N-1$}
\put(324,160){$N-2$}
\put(279,160){$N-3$}
\put(234,160){$N-4$}
\put(-30,120){$m\cosh\theta$}
\end{picture}
\caption{$l_{ab}$ is $1$ if nodes $a$ and $b$ are connected in affine $D_N$ Dynkin diagram and $0$ if not. Upper diagram is for 
${\rm SST}_{\lambda=1/N}^{(+)}$ and lower for ${\rm SST}_{\lambda=1/N}^{(-)}$ with $N\ge 4$. 
For $N=3$, only non-zero $l_{ab}$ are $l_{a,a+1}=1$ with $a=0,1,2,3$ and cyclic. For $N=2$, all $l_{ab}=0$.}
\end{figure}
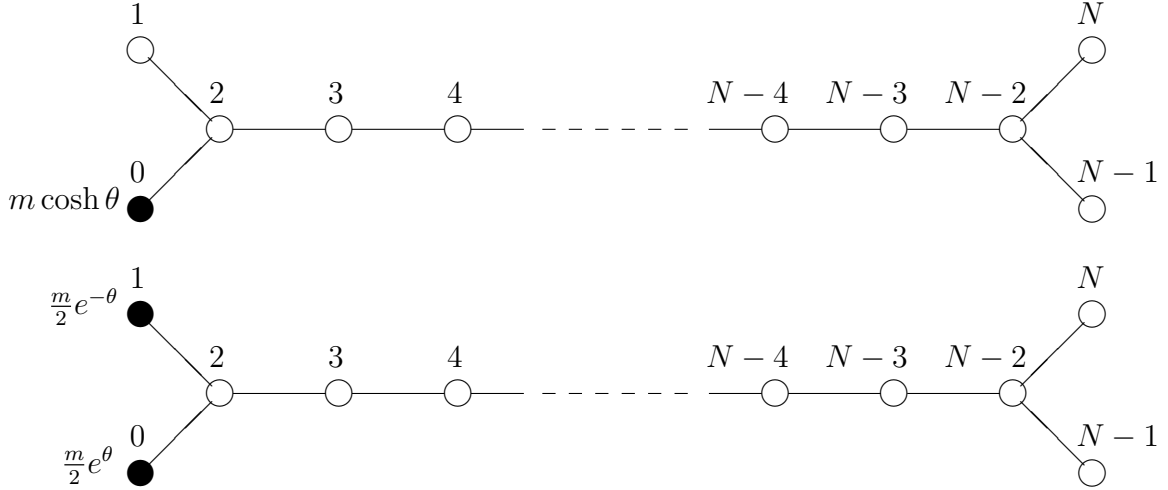
The effective central charge from the TBA system is given by
\be
c_{\rm TBA}(r)=\frac{3r}{\pi^2}\sum_{a}\int^{\infty}_{-\infty}\rho_a(\theta)\log\left(1+e^{-\varepsilon_a(\theta)}\right)d\theta.
\label{ccTBA}
\ee
Both analytic and numerical analysis have been applied for the TBA and shown that (\ref{ccTBA}) is matching with (\ref{ccSSM}) for
the special values of $\lambda=1/N$.
In the next section, we will derive NLIE which is valid for generic value of $\lambda$ in the repulsive regime.

\setcounter{equation}{0}
\section{NLIE}

\subsection{$T-Q$ system}
The TBA system can be transformed to ``$Y$-system'', 
\be
y^+ y^-&=&Y_2, \qquad y_2^+ y_2^-=Y_0Y_1Y_3, \\
y_k^+ y_k^-&=&Y_{k-1}Y_{k+1}, \quad k=3,\ldots,N-3\\
y_{N-2}^+ y_{N-2}^-&=&Y_{N-3}Y_{N-1}Y_{N}, \qquad y_{N-1}^+ y_{N-1}^-=y_{N}^+ y_{N}^-=Y_{N-2}
\label{ysystem} 
\ee
where $y_a=e^{-\varepsilon_a(\theta)}$, $Y_a=1+y_a$, and $y_a^{\pm}=y_a(\theta\pm i\pi/2)$.
For the nodes with driving terms, we can impose extra relations
\be
y_1=y, \qquad y_0=\xi y,\qquad \xi=e^{-mr\cosh\theta},\qquad&&{\rm for}\quad {\rm SST}_{\lambda=1/N}^{(+)},\\
y_0=\xi^+ y,\qquad y_1=\xi^- y,\quad \xi^{\pm}=e^{-mr\exp(\pm\theta)/2},\qquad&&{\rm for}\quad {\rm SST}_{\lambda=1/N}^{(-)}.
\ee

Next step is to identify this $Y$-system with that of the $su(2)$ system, for which we use notations
$z_k$ and $Z_k=1+z_k$ to distinguish from (\ref{ysystem}), by relating
\be
z_k^+ z_k^-&=&Z_{k-1}Z_{k+1},\qquad  z_k\equiv y_k,\qquad k=2,\ldots, N-2,\\
Z_1&=&Y_0Y_1,\qquad Z_{N-1}=Y_{N-1}Y_N.
\label{zsystem}
\ee
For this regular part we can find corresponding ``$T$-system'' (we are using the $\phi=1$ gauge)
\be
T^{+}_kT^{-}_k=1+T_{k-1}T_{k+1},\qquad k=2,\ldots,N-2.
\label{tsystem}
\ee
by using (\ref{zsystem}) and relations
\be
z_k=T_{k-1}T_{k+1},\qquad T^{+}_kT^{-}_k=Z_k,\qquad k=1,\ldots,N-1.
\label{zT}
\ee
From Eqs.(\ref{ysystem}) and (\ref{tsystem}), one can notice that $y_{N-1}=y_N=T_{N-2}$.
Furthermore, along with (\ref{zsystem}), one can find
\be
Z_{N-1}=T^{+}_{N-1}T^{-}_{N-1}=1+T_{N-2}T_{N}=Y_{N-1}Y_N=(1+y_N)^2=(1+T_{N-2})^2,
\ee
which leads to
\be
T_N=2+T_{N-2}.
\label{tcondition}
\ee

For the $su(2)$ $T$-system we can always find the corresponding Baxter $T-Q$ system \cite{baxter}
\be
T_{k+1}Q^{[k]}-T^-_{k}Q^{[k+2]}={\bar Q}^{[-k-2]},\qquad
T^-_{k}{\bar Q}^{[-k]}-T_{k-1}{\bar Q}^{[-k-2]}=Q^{[k]},
\label{TQ}
\ee
where we use a short notation $f^{[k]}(\theta)\equiv f(\theta+i\pi k/2)$.
We note that both $y_k$ and $T_k$ functions are real analytic.
Following \cite{baxter}, one can eliminate ${\bar Q}$ from the $T-Q$ system to obtain the second order difference equation
\be
Q^{++}+Q^{--}=AQ,\qquad A=\frac{T^{[-k+1]}_k+T^{[-k-1]}_{k-2}}{T^{[-k]}_{k-1}},
\ee
where the coefficient $A$ becomes independent of $k$ by using (\ref{tsystem}).
Similarly, eliminating $Q$ in (\ref{TQ}), 
\be
{\bar Q}^{++}+{\bar Q}^{--}={\bar A}{\bar Q},\qquad {\bar A}=\frac{T^{[k+3]}_k+T^{[k+1]}_{k+2}}{T^{[k+2]}_{k+1}},
\ee
where ${\bar A}$ is also $k$-independent.
Therefore, inserting $k=N$ for $A$ and $k=N-2$ for ${\bar A}$, we get
\be
A=\frac{2+T^{[-N+1]}_{N-2}+T^{[-N-1]}_{N-2}}{T^{[-N]}_{N-1}},\qquad {\bar A}=\frac{2+T^{[N-1]}_{N-2}+T^{[N+1]}_{N-2}}{T^{[N]}_{N-1}}
\ee
where the identity (\ref{tcondition}) is used.
From this we obtain 
\be
{\bar A}=A^{[2N]}\qquad\to\qquad {\bar Q}=Q^{[2N]}.
\label{qbarq}
\ee
The final relation can be analytically continued for any real value of $N$, hence $\lambda$.

\subsection{NLIE from $T-Q$ system}
We define new functions by
\be
b_k=\frac{Q^{[k+2]}T_k^-}{{\bar Q}^{[-k-2]}},\qquad B_k=1+b_k=\frac{Q^{[k]}T_{k+1}}{{\bar Q}^{[-k-2]}},
\label{BQ}
\ee
which satisfy the NLIE functional equations
\be
b_k{\bar b}_k=Z_k,\qquad B^+_k{\bar B}^-_k=Z_{k+1}.
\label{bB}
\ee
Using Fourier transform relation
\be
\widetilde{ f^{[\alpha]} }(\omega)=p^\alpha{\tilde f}(\omega),\qquad p\equiv e^{\frac{\omega\pi}{2}},
\ee
we can express relations (\ref{BQ}) in Fourier space\footnote{In this subsection we denote by $\tilde f(\omega)$
the Fourier transform of the logarithmic derivative of the function $f(\theta)$.}
\be
\tilde{b}_k&=&p^{k+2}\tilde{Q}+p^{-1}\tilde{T}_k-p^{-k-2}\tilde{\bar Q},\\
\tilde{B}_k&=&p^{k}\tilde{Q}+\tilde{T}_{k+1}-p^{-k-2}\tilde{\bar Q},
\ee
and relations in (\ref{zT}) and (\ref{bB})
\be
\tilde{T}_k=\tilde{s}\tilde{Z}_k,\qquad \tilde{T}_{k+1}=\tilde{s}\tilde{Z}_{k+1},\qquad p\tilde{B}_k+p^{-1}\tilde{\bar B}_k=\tilde{Z}_{k+1},
\qquad \tilde{b}_k+\tilde{\bar b}_k=\tilde{Z}_{k},
\ee
with 
\be
\tilde{s}=\frac{1}{p+p^{-1}}=\frac{1}{2\cosh\frac{\omega\pi}{2}}.
\ee
Now using Eq.(\ref{qbarq}) which becomes 
\be
\tilde{\bar Q}=p^{2N}\tilde{Q},
\ee
we can obtain the NLIE in the Fourier space
\be
\tilde{b}_k&=&\frac{p^{N-k-2}-p^{k+2-N}}{p^{N-k-1}-p^{k+1-N}}\ \tilde{s}\ (\tilde{B}_k
-\tilde{\bar B}_k)+p^{-1}\tilde{s}\tilde{Z}_k,\\
\tilde{\bar b}_k&=&\frac{p^{N-k-2}-p^{k+2-N}}{p^{N-k-1}-p^{k+1-N}}\ \tilde{s}\ (\tilde{\bar B}_k
-\tilde{B}_k)+p\tilde{s}\tilde{Z}_k.
\ee
These equations are valid for any integer $k$.
We choose the simplest case $k=1$ and couple the NLIE part to the remaining $Y$-functions.
The NLIE for the sausage model for the ${\rm SST}_{\lambda}^{(+)}$ is written in terms of the complex 
function $b=b_1$ and real
function $y=y_1$, $y_0=\xi y$ and the kernel $K$ which happens to be that of the sine-Gordon model
\be
\tilde{K}=\frac{p^{N-3}-p^{3-N}}{p^{N-2}-p^{2-N}}\ \tilde{s}=\frac{\sinh\left(\frac{\omega\pi(1-3\lambda)}{2\lambda}\right)}
{2\sinh\left(\frac{\omega\pi(1-2\lambda)}{2\lambda}\right)\cosh\frac{\omega\pi}{2}}.
\ee
The final set of equations in the Fourier space are
\be
\tilde{b}&=&\tilde{K}\ (\tilde{B}-\tilde{\bar B})+p^{-1}\tilde{s}\tilde{Y}_1\tilde{Y}_0,\\
\tilde{\bar b}&=&\tilde{K}\ (\tilde{\bar B}-\tilde{B})+p\tilde{s}\tilde{Y}_1\tilde{Y}_0,\\
\tilde{y}&=&p\tilde{s}\tilde{B}+p^{-1}\tilde{s}\tilde{\bar B}.
\label{Fnlie}
\ee
As usual, we move the integration contours away from the real axis by a certain amount and define new functions
\be
a=b^{[\alpha]},\qquad {\bar a}={\bar b}^{[-\alpha]},\qquad 0<\alpha<1,
\ee
which can give the hybrid-NLIE equations in the rapidity space for the ${\rm SST}_{\lambda}^{(+)}$,
\be
\log a=K\star\log(1+a)-K^{[2\alpha]}\star\log(1+{\bar a})+s^{[\alpha-1]}\star\left[\log(1+y)+\log(1+\xi y)\right],\qquad\\
\log {\bar a}=K\star\log(1+{\bar a})-K^{[-2\alpha]}\star\log(1+a)+s^{[1-\alpha]}\star\left[\log(1+y)+\log(1+\xi y)\right],\quad \ \\
\log {y}=s^{[1-\alpha]}\star\log(1+a)+s^{[\alpha-1]}\star\log(1+{\bar a}).\qquad\qquad\qquad\qquad\qquad\qquad\qquad\qquad\
\label{nlie}
\ee
Here $\star$ is a convolution defined by
$f\star g(\theta)=\int_{-\infty}^{\infty}f(\theta-\theta')g(\theta')d\theta'$.
The ground-state energy is given by 
\be
E(r)=-\frac{m}{2\pi}\int_{-\infty}^{\infty}\cosh\theta\log(1+\xi y).
\label{energy}
\ee

For the ${\rm SST}_{\lambda}^{(-)}$, the NLIE equations can be similarly written as
\be
\log a=K\star\log(1+a)-K^{[2\alpha]}\star\log(1+{\bar a})+s^{[\alpha-1]}\star\left[\log(1+\xi^{+}y)+\log(1+\xi^{-} y)\right],\ \ \ \ \\
\log {\bar a}=K\star\log(1+{\bar a})-K^{[-2\alpha]}\star\log(1+a)+s^{[1-\alpha]}\star
\left[\log(1+\xi^{+}y)+\log(1+\xi^{-} y)\right],\ \ \ \\
\log {y}=s^{[1-\alpha]}\star\log(1+a)+s^{[\alpha-1]}\star\log(1+{\bar a}).\qquad\qquad\qquad\qquad\qquad\qquad\qquad\qquad\quad\ \ 
\label{nlieless}
\ee
along with the ground-state energy given by 
\be
E(r)=-\frac{m}{4\pi}\int_{-\infty}^{\infty}\left[e^{\theta}\log(1+\xi^{+}y)+e^{-\theta}\log(1+\xi^{-} y)\right].
\label{energyless}
\ee

These are our proposal for the NLIE equations of the sausage model with generic coupling $0\le\lambda<1/2$.

\subsection{IR limit: Next-to-leading order vacuum correction}
As a check for the NLIE of the ${\rm SST}_{\lambda}^{(+)}$, we consider the IR limit, $mr\gg 1$. 
In this limit, the variables can be expanded as
\be
a=z(1+w+\ldots),\qquad y=h(1+u+\ldots),
\ee
where $z,h$ are the leading coefficients which are finite but $\theta$-independent, and $u,w$ 
are next-to-leading order of ${\cal O}(e^{-mr})$.
Inserting into the NLIE equations, it is easy to find that $z=2$ and $h=3$.
Then the NLIE is linearized as follows:
\be
w&=&\frac{2}{3}K\star w-\frac{2}{3}K^{[2\alpha]}\star{\bar w}+s^{[\alpha-1]}\star\left(3\xi+\frac{3}{4}u\right),\\
{\bar w}&=&\frac{2}{3}K\star{\bar w}-\frac{2}{3}K^{[2\alpha]}\star{w}+s^{[1-\alpha]}\star\left(3\xi+\frac{3}{4}u\right),\\
u&=&\frac{2}{3}s^{[1-\alpha]}\star w+\frac{2}{3}s^{[\alpha-1]}\star{\bar w}.
\ee
These linearized equations can be readily solved by Fourier transforms,
\be
\tilde{u}=\frac{1}{3}\tilde{\xi}\tilde{\varphi},\qquad
\tilde{\varphi}=8\ \frac{\sinh\left[\pi\omega\left(\frac{1}{2\lambda}-1\right)\right]}{\sinh\frac{\pi\omega}{2\lambda}}
-4\ \frac{\sinh\left[\pi\omega\left(\frac{1}{2\lambda}-2\right)\right]}{\sinh\frac{\pi\omega}{2\lambda}}.
\label{linearsol}
\ee
The energy in (\ref{energy}) can be also expanded as
\be
E&=&E^{(1)}+E_1^{(2)}+E_2^{(2)}+{\cal O}(e^{-3mr}),\\
E^{(1)}&=&-\frac{e_1m}{2\pi}\int_{-\infty}^{\infty}\cosh\theta\ e^{-mr\cosh\theta}\ d\theta,\\
E_1^{(2)}&=&\frac{e_2m}{4\pi}\int_{-\infty}^{\infty}\cosh\theta\ e^{-2mr\cosh\theta}\ d\theta,\\
E_2^{(2)}&=&-\frac{m}{2\pi}\int_{-\infty}^{\infty}d\theta\cosh\theta\ e^{-mr\cosh\theta}\ \int_{-\infty}^{\infty}d\theta'\varphi(\theta-\theta')
\ e^{-mr\cosh\theta'}\ d\theta',
\ee
with $e_1=3,\ e_2=9$ and $\varphi(\theta)$ is the Fourier transform of $\tilde{\varphi}$ in (\ref{linearsol}).

From the L\"uscher expansion we get the same formulae with $e_1=n,\ e_2=n^2$, where $n$ is the number of particles and 
\be
\varphi(\theta)=\frac{1}{2\pi i}\frac{d}{d\theta}\log {\rm det} S_\lambda^{(+)}(\theta),
\ee
where  $S_\lambda^{(+)}$ is the S-matrix (\ref{SSTpp})-(\ref{SSTp}).
One can check that the expansion is consistent with the triplet spectrum for the ${\rm SST}_{\lambda}^{(+)}$ and 
the $S$-matrix.

\subsection{UV limit}
We have checked numerically that the NLIE system matches with the TBA system accurately for the values of $1/\lambda$=integer in 
the limit $mr\ll 1$.
To understand the UV limit in more details, we generate the effective central charge from the NLIE system for generic $\lambda$ 
(and from the TBA for integer values of $1/\lambda$) and compare with analysis based on the ${\rm SSM}^{(\theta)}_\nu$ action.
Analysis based on the zero-mode dynamics in \cite{FOZ}, which leads to (\ref{ccSSM}), is 
\be
c(r)=2-\frac{4\pi}{\nu}\ \frac{3\pi^2}{2(\log(mr)+\delta)^2},
\label{newquant}
\ee
and we made a quadratic polynomial fit to the data points $1/(2-c(r))$ in the variable $\log (mr)$.
\begin{figure}[t]
\begin{center}
\includegraphics[width=450pt, height=250pt]{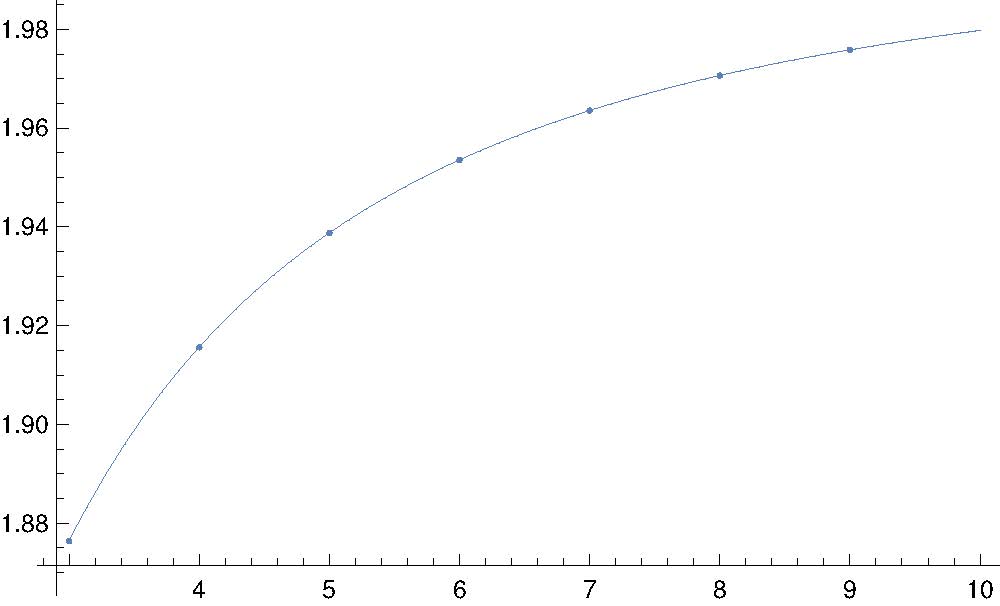}
\end{center} 
\caption{Effective central charge $c(r)$ vs. $-\log_{10} (mr)$: 
NLIE (dots) and quadratic fitting (curve) for $\lambda=1/3$.  }
\end{figure}
In Fig.2, we have plotted the effective central charge $c(r)$ vs. $-\log_{10}(mr)$ where dots are those from numerical 
solutions of the NLIE and
the curve is the fitted quadratic polynomial in Eq.(\ref{newquant}) for $\lambda=1/3$.
It shows an excellent agreement between the NLIE and reflection relation in the UV limit.

Furthermore, from the coefficient of the quadratic term, one can find exact $\nu-\lambda$ relation for generic
value of $\lambda$.
From numerical analysis summarized in Table 1, we conclude that exact $\nu-\lambda$ relation  should be
\be
\frac{\nu}{4\pi}=\frac{\lambda}{1-2\lambda}.
\label{exactnulam}
\ee
\begin{table}[ht]
\centering
\begin{tabular}{|c |c |c|}
\hline
& &  \\
$\lambda$&$\frac{\nu}{4\pi}$ (numeric)&$\frac{\nu}{4\pi}$ by (\ref{exactnulam}) \\[3ex]
\hline
& &  \\
1/2.7&1.4289&1.4286\\[2ex]
\hline
& &  \\
1/2.9 &1.1113&1.1111 \\[2ex]
\hline
& &  \\
1/3.2&0.8333&0.8333 \\[2ex]
\hline
& &  \\
1/4.5&0.3992&0.4000 \\[2ex]
\hline
& &  \\
1/5.5&0.2845&0.2857\\[2ex]
\hline
\end{tabular}
\caption{Quadratic fitting data in the UV limit}
\label{table:nonlin}
\end{table}
In Appendix A, we have derived the effective central charge in the UV limit analytically by utilizing a method used in the study of the sinh-Gordon model
in this limit \cite{ZamLio,Ahn:1999un,Fring:1999mn} and proved (\ref{exactnulam}) analytically.
This exact $\nu-\lambda$ relation (\ref{exactnulam}) is consistent with the TBA result valid for integer values of $1/\lambda$ \cite{FOZ}.

\setcounter{equation}{0}
\section{Conclusion}
The sausage model is guiding how to generalize integrable NLS models.
In this paper, we have proposed the NLIE for the model which are valid for generic values of $\lambda$
in the repulsive domain of $0\le\lambda<1/2$.
We have analyzed both IR and UV behaviours of the NLIE to establish direct connections with the 
$S$-matrix and an exact relation between $\lambda$ and $\nu$.

A number of releted issues need further studies.
It will be interesting to elaborate more on the zero-mode dynamics to that of reflection amplitude of the sine-Liouville
theory \cite{sineLiou}.
Another challenge is to extend either TBA or NLIE to the sausage model in the  ``attractive'' domain $\lambda>1/2$. 
Understanding these cases will certainly help constructing NLIEs for the deformed AdS/CFT systems, which
gets a lot of attention recently.

\section*{Acknowledgements}
This work was supported by the National Research Foundation of Korea (NRF) grant
(NRF-2016R1D1A1B02007258) (CA), by the Hungarian National Science Fund OTKA (under K116505) (JB), and 
by Commission IV (Theory) of I.N.F.N. under the grant GAST (FR).
FR thanks D. Fioravanti, A. Bonini and N. Vernazza for useful discussions.

\vskip 2cm
\appendix
\renewcommand{\theequation}{A.\arabic{equation}}
\setcounter{equation}{0}
\section{Appendix: UV expansion of the sausage NLIE}
\subsection{NLIE setup}

Introducing the notation
\begin{equation}
y(\theta)={\rm e}^{-\varepsilon_1(\theta)},\qquad
\xi(\theta)y(\theta)={\rm e}^{-\varepsilon_0(\theta)},\qquad
a(\theta)={\rm e}^{-\varepsilon_2(\theta)},\qquad
\bar a(\theta)={\rm e}^{-\varepsilon_3(\theta)},
\end{equation}
where
\begin{equation}
\varepsilon_1(\theta)\ \ {\rm and}\ \ 
\varepsilon_0(\theta)=\varepsilon_1(\theta)+mr\cosh\theta\ \  {\rm are\ real,\ \ \  and}\ \ 
[\varepsilon_2(\theta)]^*=\varepsilon_3(\theta),
\end{equation}
we can rewrite the sausage NLIE in the \lq\lq TBA-like'' form
\begin{equation}
\delta_{a0}\,mr\cosh\theta=\varepsilon_a(\theta)+\frac{1}{2\pi}\sum_b\int_{-\infty}^\infty
{\rm d}\theta^\prime\Psi_{ab}(\theta-\theta^\prime)L_b(\theta^\prime),\qquad
L_a=\ln\left(1+{\rm e}^{-\varepsilon_a}\right).
\label{NLIE0}
\end{equation}
Here the kernel matrix is
\begin{equation}
\Psi=\begin{pmatrix}
0&0&s^{[1-\alpha]}&s^{[\alpha-1]}\\
0&0&s^{[1-\alpha]}&s^{[\alpha-1]}\\
s^{[\alpha-1]}&s^{[\alpha-1]}&K&-K^{[2\alpha]}\\
s^{[1-\alpha]}&s^{[1-\alpha]}&-K^{[-2\alpha]}&K
\end{pmatrix}.
\end{equation}
$K$ and $s$ are even, real analytic functions and (for real $\theta$)
this implies the relations
\begin{equation}
[\Psi_{ab}(\theta)]^*=\Psi_{ba}(\theta)=\Psi_{ab}(-\theta),
\end{equation}
and in Fourier space the relations
\begin{equation}
[\tilde\Psi_{ab}(\omega)]^*=\tilde\Psi_{ab}(\omega),\qquad
\tilde\Psi_{ab}(-\omega)=\tilde\Psi_{ba}(\omega).
\end{equation}
We write the Taylor expansion of the Fourier kernels as
\begin{equation}
\tilde\Psi_{ab}(\omega)=\sum_{n=0}^\infty(-i)^n\tilde\Psi_{ab,n}\omega^n.
\end{equation}
The Taylor coefficients satisfy the symmetry relations
\begin{equation}
\tilde\Psi^*_{ab,n}=(-1)^n\tilde\Psi_{ab,n}\qquad
\tilde\Psi_{ab,n}=(-1)^n\tilde\Psi_{ba,n}.
\end{equation}
The NLIE equations (\ref{NLIE0}) are consistent with the symmetry relations
\begin{equation}
L_2^*(\theta)=L_3(\theta)=L_2(-\theta).
\end{equation}
Furthermore, $L_0(\theta)$ and $L_1(\theta)$ must be real and even.
For later purposes we calculate the first few Taylor coefficients:
\begin{equation}
\tilde\Psi_{ab,0}=\begin{pmatrix}
0&0&1/2&1/2\\
0&0&1/2&1/2\\
1/2&1/2&p/2&-p/2\\
1/2&1/2&-p/2&p/2
\end{pmatrix},\qquad\quad p=\frac{N-3}{N-2},
\end{equation}
\begin{equation}
\tilde\Psi_{ab,1}=\begin{pmatrix}
0&0&h_1&-h_1\\
0&0&h_1&-h_1\\
-h_1&-h_1&0&-l_1\\
h_1&h_1&l_1&0
\end{pmatrix},\qquad\quad h_1=\frac{i\pi(1-\alpha)}{4},\quad l_1=\frac{i\alpha\pi p}{2},
\end{equation}
\begin{equation}
\tilde\Psi_{ab,2}=\begin{pmatrix}
0&0&h_2&h_2\\
0&0&h_2&h_2\\
h_2&h_2&q_2&l_2\\
h_2&h_2&l_2&q_2
\end{pmatrix},\quad h_2=\frac{\pi^2(2\alpha-\alpha^2)}{16},
\quad q_2=\frac{p(N-1)\pi^2}{24},\quad l_2=\frac{p\alpha^2\pi^2}{4}-q_2.
\end{equation}

Using the Fourier coefficients we can formally rewrite the NLIE integral equations in the
form of infinite order differential equations:
\begin{equation}
\delta_{a0}\,mr\cosh\theta=\varepsilon_a(\theta)+\sum_b\sum_{n=0}^\infty \tilde\Psi_{ab,n}
L_b^{(n)}(\theta),
\label{NLIE1}
\end{equation}
where 
\begin{equation}
L_a^{(n)}(\theta)=\frac{{\rm d}^n}{{\rm d}\theta^n}L_a(\theta).
\end{equation}
It is convenient to write (\ref{NLIE1}) in terms of the functions $L_a$ only:
\begin{equation}
\delta_{a0}\,mr\cosh\theta+\sum_b{\cal M}_{ab}L_b(\theta)
+\ln\left(1-{\rm e}^{-L_a(\theta)}\right)
=\sum_b\sum_{n=1}^\infty \tilde\Psi_{ab,n}
L_b^{(n)}(\theta).
\label{NLIE2}
\end{equation}
Here we introduced the matrix
\begin{equation}
{\cal M}_{ab}=\delta_{ab}-\tilde\Psi_{ab,0}.
\end{equation}
Note that this matrix has a zero mode:
\begin{equation}
\sum_b{\cal M}_{ab}=\sum_a{\cal M}_{ab}=0.
\end{equation}

The effective central charge is given by
\begin{equation}
c(r)=\frac{6mr}{\pi^2}\int_0^\infty{\rm d}\theta\cosh\theta L_0(\theta).
\end{equation}
In the UV limit, $r\to0$, the $\cosh\theta$ factor can be approximated by ${\rm e}^\theta/2$:
\begin{equation}
c(r)\approx\frac{3mr}{\pi^2}\int_0^\infty{\rm d}\theta\,{\rm e}^\theta L_0(\theta).
\end{equation}
Here and in the following the meaning of the symbol $\approx$ is that the error is 
power-like, O$(r^\gamma)$, for $r\to0$. Introducing the variable
\begin{equation}
x=\ln\frac{2}{mr},
\end{equation}
for $x\to\infty$ the error is exponentially small, O$({\rm e}^{-\gamma x})$.

\subsection{Zamolodchikov trick}

Zamolodchikov introduced the integral
\begin{equation}
\tilde c(r,y)=\frac{3mr}{\pi^2}\int_y^\infty{\rm d}\theta\,{\rm e}^\theta L_0(\theta).
\end{equation}
The same function is defined by the relations
\begin{equation}
\frac{\partial\tilde c(r,y)}{\partial y}=-\frac{6}{\pi^2}\,{\rm e}^{y-x}L_0(y),\qquad
\tilde c(r,\infty)=0.
\end{equation}
$\tilde c$ is a useful function, because, as it is easy to see, in the central region
\begin{equation}
-Ax<y<Ax\qquad {\rm for\ any\ }0<A<1
\end{equation}
\begin{equation}
c(r)\approx\tilde c(r,y).
\end{equation}
Let us assume that $y$ is in the central region or larger: $y>-Ax$. In this region
(\ref{NLIE2}) can be approximated by
\begin{equation}
\delta_{a0}\,{\rm e}^{y-x}+\sum_b{\cal M}_{ab}L_b(y)
+\ln\left(1-{\rm e}^{-L_a(y)}\right)
\approx\sum_b\sum_{n=1}^\infty \tilde\Psi_{ab,n}
L_b^{(n)}(y).
\label{NLIE3}
\end{equation}
Following an analogous construction in the sinh-Gordon model we define
\begin{equation}
\begin{split}
\tilde{\tilde c}(r,y)=\frac{3}{\pi^2}\Bigg\{\sum_{n=2}^\infty\sum_{a,b}&\tilde\Psi_{ab,n}
\sum_{k=1}^{n-1}(-1)^{k+1}L_a^{(k)}(y)L_b^{(n-k)}(y)-2L_0(y){\rm e}^{y-x}\\
&-\sum_{a,b}L_a(y){\cal M}_{ab}L_b(y)-2\sum_a{\rm Li}_2\left({\rm e}^{-L_a(y)}\right)
\Bigg\},
\end{split}
\end{equation}
where ${\rm Li}_2$ is the dilogarithm function
\begin{equation}
{\rm Li}_2(z)=\sum_{n=1}^\infty\frac{z^n}{n^2}.
\end{equation}
It is related to the Rogers dilogarithm ${\cal L}$ by
\begin{equation}
{\cal L}(z)={\rm Li}_2(z)+\frac{1}{2}\ln z\ln(1-z).
\end{equation}

Using (\ref{NLIE3}) it is easy to show that $\tilde{\tilde c}$ satisfies
\begin{equation}
\frac{\partial\tilde{\tilde c}(r,y)}{\partial y}
\approx -\frac{6}{\pi^2}\,{\rm e}^{y-x}L_0(y),
\end{equation}
which means that the functions $\tilde c$ and $\tilde{\tilde c}$ differ by a constant only.
To calculate this constant we consider $\tilde{\tilde c}(r,\infty)$. In the limit $y\to\infty$
\begin{equation}
L_0(y)\to0
\end{equation}
and the other three $L$-functions go to constant values:
\begin{equation}
L_\mu(y)\to \hat L_\mu,\quad \mu=1,2,3,
\end{equation}
where
\begin{equation}
\hat L_\mu=\ln(1+\hat x_\mu),\qquad \hat x_\mu={\rm e}^{-\varepsilon_\mu(\infty)}.
\end{equation}
Using the symmetry properties we write
\begin{equation}
\hat x_1=h>0\quad{\rm real},\qquad \hat x_2=z,\quad \hat x_3=z^*.
\end{equation}
In terms of these variables the asymptotic NLIE equations
\begin{equation}
\sum_\nu{\cal M}_{\mu\nu}\hat L_\nu+\ln\left(1-{\rm e}^{-\hat L_\mu}\right)=0
\end{equation}
can be written
\begin{equation}
h=\vert1+z\vert,\qquad
z=\sqrt{1+h}\exp\{ip\,{\rm arg}(1+z)\}.
\end{equation}
Since
\begin{equation}
\vert{\rm arg}(1+z)\vert\leq\vert{\rm arg}(z)\vert,
\end{equation}
for $\vert p\vert<1$ ($N>5/2$) $z$ must be real and positive and the only solution is
\begin{equation}
z=2,\qquad h=3.
\end{equation}
We can now calculate
\begin{equation}
\begin{split}
\tilde{\tilde c}(r,\infty)&=\frac{3}{\pi^2}\left\{-\sum_{\mu,\nu}
\hat L_\mu{\cal M}_{\mu\nu}\hat L_\nu-\frac{\pi^2}{3}-2\sum_\mu{\cal L}
\left({\rm e}^{-\hat L_\mu}\right)-\sum_\mu\hat L_\mu\ln\left(1-{\rm e}^{-\hat L_\mu}\right)
\right\}\\
&=-1-\frac{6}{\pi^2}\sum_\mu{\cal L}\left({\rm e}^{-\hat L_\mu}\right)
=-1-\frac{6}{\pi^2}\left\{{\cal L}(1/4)+2{\cal L}(1/3)\right\}=-2.
\end{split}
\end{equation}
In the second line we used the identity
\begin{equation}
\sum_{k=2}^n{\cal L}\left(\frac{1}{k^2}\right)+2{\cal L}\left(\frac{1}{n+1}\right)
=\frac{\pi^2}{6}
\end{equation}
for $n=2$. We conclude
\begin{equation}
\tilde c(r,y)\approx 2+\tilde{\tilde c}(r,y).
\end{equation}

In the central region we thus have
\begin{equation}
\begin{split}
c(r)\approx 2+\frac{6}{\pi^2}\Bigg\{\frac{1}{2}\sum_{n=2}^\infty\sum_{a,b}&\tilde\Psi_{ab,n}
\sum_{k=1}^{n-1}(-1)^{k+1}L_a^{(k)}(y)L_b^{(n-k)}(y)\\
&-\frac{1}{2}\sum_{a,b}L_a(y){\cal M}_{ab}L_b(y)
-\sum_a{\rm Li}_2\left({\rm e}^{-L_a(y)}\right)
\Bigg\}.
\end{split}
\label{central}
\end{equation}
In the same region we can neglect the mass term and simplify the NLIE equations:
\begin{equation}
\sum_b{\cal M}_{ab}L_b(y)
+\ln\left(1-{\rm e}^{-L_a(y)}\right)
\approx\sum_b\sum_{n=1}^\infty \tilde\Psi_{ab,n}
L_b^{(n)}(y).
\label{NLIE4}
\end{equation}

\subsection{No plateau solution}

Usually in the UV limit $x\to\infty$ a long plateau is formed, in the central region
the $L$-functions are approximately constant and satisfy the constant version of
(\ref{NLIE4}):
\begin{equation}
\sum_b{\cal M}_{ab}L_b
+\ln\left(1-{\rm e}^{-L_a}\right)=0.
\end{equation}
Introducing
\begin{equation}
x_a={\rm e}^{-\varepsilon_a(0)},\qquad L_a=\ln(1+x_a)
\end{equation}
the symmetry properties imply
\begin{equation}
x_0=x_1=h>0\quad{\rm real},\qquad x_2=x_3=z \quad{\rm real} 
\end{equation}
and the constant NLIE becomes
\begin{equation}
h=\vert1+z\vert,\qquad z=(1+h)\exp\{ip\,{\rm arg}(1+z)\}.
\end{equation}
For $\vert p\vert<1$ the only possibility is
\begin{equation}
{\rm arg}(z)={\rm arg}(1+z)=0 \qquad (z>0).
\end{equation}
The constant NLIE is reduced to the contradictory pair of equations
\begin{equation}
h=1+z,\qquad z=1+h.
\end{equation}
Thus there is no plateau solution and we conclude that, similarly to what happens in the
sinh-Gordon model,
\begin{equation}
\varepsilon_a(0)\to-\infty\qquad (x\to\infty).
\end{equation}

\subsection{Zamolodchikov Ansatz}

We now introduce a \lq\lq coupling constant'' $g$ which goes to zero as $x\to\infty$
and (in the central region) expand the $L$-functions perturbatively as
\begin{equation}
L_a(\theta)=W_a\ln\left(\frac{1}{g^2}\right)+\ell_{a,0}(g\theta)+g\ell_{a,1}(g\theta)
+g^2\ell_{a,2}(g\theta)+\dots
\end{equation}
Putting this expansion into (\ref{NLIE4}) the O$(\ln g^2)$ (divergent) term gives
\begin{equation}
\sum_b{\cal M}_{ab}W_b=0,
\end{equation}
so $W_a$ must be proportional to the zero mode,
\begin{equation}
W_a=W.
\end{equation}
We choose $W=1$ so that the \lq\lq potential'' term
\begin{equation}
\ln\left(1-{\rm e}^{-L_a(y)}\right)
\end{equation}
can also be expanded in integer powers of the coupling and we have a consistent 
perturbation theory that can be solved order by order in $g$. At O$(1)$ we have
\begin{equation}
\sum_b{\cal M}_{ab}\ell_{b,0}(\zeta)=0.
\end{equation}
It follows that $\ell_{a,0}$ is also proportional to the zero mode:
\begin{equation}
\ell_{a,0}(\zeta)=\ell_0(\zeta).
\end{equation}
The (so far) undetermined function $\ell_0$ must be real and even. The O$(g)$ equation is
\begin{equation}
\sum_b{\cal M}_{ab}\ell_{b,1}(\zeta)=\sum_b\tilde\Psi_{ab,1}\ell_0^\prime(\zeta).
\end{equation}
We introduce
\begin{equation}
\eta_a=\sum_b\tilde\Psi_{ab,1}=\frac{i\pi}{2}(\alpha-1-\alpha p)
\begin{pmatrix}
0\\0\\1\\-1\end{pmatrix}
\end{equation}
and define $Y_a$ as the solution of the linear equation
\begin{equation}
\sum_b{\cal M}_{ab}Y_b=\eta_a
\end{equation}
with the auxilliary condition
\begin{equation}
Y_0=Y_1=0.
\end{equation}
(This extra condition is necessary to make the solution unique since the matrix
${\cal M}$ is degenerate.) We find
\begin{equation}
Y_a=q
\begin{pmatrix}
0\\0\\1\\-1\end{pmatrix},
\qquad q=\frac{(\alpha-1-\alpha p)i\pi}{2(1-p)}.
\end{equation}
The general solution of the O$(g)$ problem is
\begin{equation}
\ell_{a,1}(\zeta)=Y_a\ell_0^\prime(\zeta)+\ell_1(\zeta),
\end{equation}
where the undetermined function $\ell_1$ must be real and even.

At O$(g^2)$ we have
\begin{equation}
\sum_b{\cal M}_{ab}\ell_{b,2}(\zeta)-{\rm e}^{-\ell_0(\zeta)}
=\sum_b\tilde\Psi_{ab,1}\ell_{b,1}^\prime(\zeta)
+\sum_b\tilde\Psi_{ab,2}\ell_0^{\prime\prime}(\zeta).
\end{equation}
We will now use the consistency of this system to determine $\ell_0$. Summing over $a$
we have
\begin{equation}
-4{\rm e}^{-\ell_0(\zeta)}
=-\sum_a\eta_a\ell_{a,1}^\prime(\zeta)
+\sum_{a,b}\tilde\Psi_{ab,2}\ell_0^{\prime\prime}(\zeta)
=2B\ell_0^{\prime\prime}(\zeta),
\end{equation}
where
\begin{equation}
B=\frac{1}{2}\sum_{a,b}\tilde\Psi_{ab,2}-\frac{1}{2}\sum_a\eta_aY_a=\frac{\pi^2(N-2)}{4}.
\end{equation}
Note that the constant $B$ is $\alpha$-independent.

The second order differential equation satisfied by $\ell_0$ is of the same form as
in the sinh-Gordon model and its even solution is unique (up to rescaling the coupling):
\begin{equation}
\ell_0(\zeta)=\ln\frac{\cos^2(\zeta)}{B}.
\end{equation}
From this solution we see that in the central region ${\rm e}^{L_a(\theta)}$ is 
everywhere large,
of the order $1/g^2$:
\begin{equation}
{\rm e}^{L_a(\theta)}=\frac{1}{g^2}\left(\frac{\cos^2(g\theta)}{B}+\dots\right)
\end{equation}
But there is no reason why at the boundary of the central region, at $\theta=x$,
${\rm e}^{L_a(\theta)}$ should be large. We require it is O$(1)$ and this fixes
the relation between the coupling and $x$:
\begin{equation}
g\sim\frac{\pi}{2x}.
\end{equation}
More precisely, the coupling must have a large $x$ expansion of the form
\begin{equation}
g=\frac{\pi}{2}\left(\frac{1}{x}+\frac{g_2}{x^2}+\dots\right).
\end{equation}
It is not possible to determine the higher terms in this expansion with the present 
method, only the leading term is fixed.

Using the perturbative solution in (\ref{central})
we can calculate the perturbative expansion of the central charge: 
\begin{equation}
c(r)\approx 2+g^2\Delta_2+\dots,
\end{equation}
where
\begin{equation}
\Delta_2=\frac{6}{\pi^2}\left\{B\ell_0^{\prime2}-4{\rm e}^{-\ell_0}\right\}=
-\frac{24B}{\pi^2}=-6(N-2).
\end{equation}
The final result is
\begin{equation}
c(r)\approx 2-\frac{3\pi^2}{2}\frac{N-2}{x^2}+\dots
\end{equation}

%\end{appendix}


\begin{thebibliography}{99}
\bibitem{Malda} J. Maldacena, ``The large N limit of superconformal
field theories and supergravity,''  Adv. Theor. Math. Phys. {\bf 2},
231 (1998) [{arXiv:hep-th/9711200}].

\bibitem{review} N. Beisert, C. Ahn, et.al., ``Review of AdS/CFT
Integrability: An Overview,'' Lett. Math. Phys. {\bf 99}, 3 (2012)
[arXiv:hep-th/1012.3982v5].

\bibitem{tba} Al. B. Zamolodchikov, Nucl. Phys. {\bf B342} (1990) 695.

\bibitem{nlie1} A. Kl\"umper, M.T. Batchelor, J. Phys. {\bf A23} (1990) L189;
A. Kl\"umper, P.A. Pearce, J. Stat. Phys. {\bf 64} (1991) 13;
A. Kl\"umper, M.T. Batchelor, P.A. Pearce, J. Phys. {\bf A24} (1991) 3111.
\bibitem{nlie2} C. Destri, H. de Vega, Phys. Rev. Lett. {\bf 69} (1992)
2313; Nucl. Phys. {\bf B438} (1995) 413;
Nucl. Phys. {\bf B504} (1997) 621.
\bibitem{nlie3} D. Fioravanti, A. Mariottini, E. Quattrini, F. Ravanini, Phys. Lett. {\bf B390} (1997) 243;
G. Feverati, F. Ravanini, G. Tak\'acs, Phys. Lett. {\bf B430} (1998) 264, 
Nucl. Phys. {\bf B540} (1999) 543, Phys. Lett. {\bf B444} (1998) 442.

\bibitem{BalHeg} J\'anos Balog, \'Arp\'ad Hegedus,  Nucl. Phys. {\bf B829} (2010) 425.

\bibitem{adsnlie} N. Gromov, V. Kazakov, P. Vieira, JHEP {\bf 0912} (2009) 060.

\bibitem{Gromov:2011cx}
  N.~Gromov, V.~Kazakov, S.~Leurent and D.~Volin,
  %``Solving the AdS/CFT Y-system,''
  JHEP {\bf 1207} (2012) 023
  doi:10.1007/JHEP07(2012)023
  [arXiv:1110.0562 [hep-th]].
  
\bibitem{FOZ} V. A. Fateev, E. Onofri, and Al. B. Zamolodchikov, ``Integrable deformations of the $0(3)$ sigma model. The sausage model,''
Nucl. Phys. {\bf B406} (1993) [FS] 521.

\bibitem{adsdeform1} S. Frolov, JHEP 0505 (2005) 069;
D.V. Bykov, S. Frolov, JHEP 0807 (2008) 071.

\bibitem{adsdeform2} F. Delduc, M. Magro, B. Vicedo,
``An integrable deformation of the $AdS_5 \times S^5$ superstring action,''
Phys. Rev. Lett. {\bf 112}, 051601 (2014)
arXiv:1309.5850 [hep-th]; ``On classical q-deformations of integrable sigma-models,''
JHEP {\bf 1311}, 192 (2013)  [arXiv:1308.3581 [hep-th]].

\bibitem{adsdeform3} C. Klimcik, ``Yang-Baxter sigma models and dS/AdS T duality,'' JHEP {\bf 0212}, 051 (2002)
[hep-th/0210095]; ``On integrability of the Yang-Baxter sigma-model,'' J. Math. Phys. {\bf 50},
043508 (2009) [arXiv:0802.3518 [hep-th]].

\bibitem{adsdeform4} B. Hoare, R. Roiban, and A. A. Tseytlin, ``On deformations of $AdS_n \times S^n$ supercosets,'' JHEP {\bf 1406}, 002 (2014)[arXiv:1403.5517 [hep-th]].


\bibitem{BeiKor} N. Beisert, P. Koroteev, ``Quantum Deformations of the One-Dimensional Hubbard
Model," J. Phys. {\bf A41} (2008) 255204.

\bibitem{ZamZam1} A. B. Zamolodchikov and Al. B. Zamolodchikov, Ann. Phys. {\bf 120} (1979) 253.
\bibitem{ZamZam2} A. B. Zamolodchikov and Al. B. Zamolodchikov, Nucl. Phys. {\bf B379} (1992) 602.

\bibitem{baxter} P. Wiegmann, ``Bethe ansatz and classical Hirota equation,''
Int. J. Mod. Phys. {\bf B11} (1997) 75 [cond-mat/9610132].

\bibitem{ZamLio} Al. B. Zamolodchikov, ``Resonance factorized scattering and roaming trajectories,'' preprint
ENS-LPS-335 (1991), J. Phys. {\bf A39} (2006) 12847.

\bibitem{Ahn:1999un}
  C.~Ahn, C.~Kim and C.~Rim,
%``Hidden relation between reflection amplitudes and thermodynamic Bethe ansatz,''
  Nucl.\ Phys.\ B {\bf 556} (1999) 505
  doi:10.1016/S0550-3213(99)00405-8
  [hep-th/9903134].

\bibitem{Fring:1999mn}
  A.~Fring, C.~Korff and B.~J.~Schulz,
  %``The Ultraviolet behavior of integrable quantum field theories, affine Toda field theory,''
  Nucl.\ Phys.\ B {\bf 549} (1999) 579
  doi:10.1016/S0550-3213(99)00216-3
  [hep-th/9902011].


\bibitem{sineLiou} V. Fateev, A. Zamolodchikov and Al. Zamolodchikov, unpublished.



\end{thebibliography}
\end{document}